\documentclass[12pt]{article}
\begin{document}
\title{Signature of the Planck Scale}
\author{B.G.Sidharth\\
Dipartimento di Matematica e Informatica,\\ Universita di Udine, Via
delle Scienze 206,\\ 33100 Udine Italy}
\date{}
\maketitle
\begin{abstract}
A fundamental spacetime scale in the universe leads to
noncommutative spacetime and thence to a modified energy - momentum
dispersion relation or equivalently to a modification of Lorentz
symmetry as shown by the author and others. This latter
consideration has also been used by some scholars though based on
purely phenomenological models that have been suggested by the
observation of Ultra High Energy Cosmic Rays. On the other hand a
parallel development has been the proposal of a small but non zero
photon mass by some scholars including the author, such a mass being
within experimentally allowable limits. This too leads to a small
violation of Lorentz symmetry observable in principle in very high
energy gamma rays, as in fact is claimed. We show in this paper that
the latter mechanism in fact follows from the former, thus unifying
two apparently different approaches. We also examine this scenario
in the context of Fermions and show some interesting results.
\end{abstract}
\section{Introduction}
Glashow and several others have considered a small modification of
the velocity of light or of Lorentz symmetry based on hints from
observation of Ultra High Energy cosmic Rays
\cite{bgsfpl,tduniv,ijtp1}. However the High Energy Gamma Ray data
available at that time gave rather large limits for the variation of
the speed of light or the mass of the photon, for example
$\frac{\Delta c}{c} \sim 10^{-21}$ or $m_\gamma$, the photon mass
less than $10^{-44}gms$ (Cf.ref.\cite{shaefer}). Independently the
author had come to a similar conclusion though from a purely
theoretical point of view namely the existence of a fundamental
spacetime scale, the Planck scale resulting in a noncommutative
spacetime geometry \cite{bgscsf,cu,bgsijmpe,uof}. Such a conclusion
about a violation of Lorentz symmetry also comes from a different
and surprising angle as deduced by the author (and a few others) -
that of a non zero photon mass proposed very early on by De Broglie,
amongst others
\cite{bgsmp,tduniv,debroglievig,debroglie1,debroglie2}. It may be
mentioned that these latter values of the author are well within the
tighter and
improved experimental limits.\\
We will now argue that indeed the noncommutative spacetime approach,
suggested by recent Quantum Gravity theories, leads directly to the
same photon mass, thus unifying two approaches, that of a
fundamental spacetime scale and that of the photon mass. We will
also
discuss the observational support for all this.\\
It should also be mentioned that in the author's work spacetime has
an underpinning of Planck scale oscillators in a background Dark
Energy - the ZPF, and this indeed had led in 1997 to the prediction
of an accelerating universe with a small cosmological constant,
besides several other consistent with observation results
\cite{bgsijmpa98,ijtp2,cu,uof}. Further it has been shown by the
author and others that the Planck oscillator scale is a minimum -
smaller scales are physically meaningless
\cite{tduniv,ijmpe,cer1,cer2}.
\section{The Modified Dispersion Relation}
To see this in greater detail, we note that, given a minimum length
$l$, we saw that the usual commutation relations get modified and
now become
\begin{equation}
[x,p] = \hbar' = \hbar [1 + \left(\frac{l}{\hbar}\right)^2 p^2]\,
etc\label{5He2}
\end{equation}
(Cf. also ref.\cite{bgsust}). (\ref{5He2}) shows that effectively
$\hbar$ is replaced by $\hbar'$. So, in units, $\hbar = 1 = c$,
$$E = [m^2 + p^2 (1 + l^2 p^2)^{-2}]^{\frac{1}{2}}$$
or, the energy-momentum relation leading to the Klein-Gordon
Hamiltonian is given by,
\begin{equation}
E^2 = m^2 + p^2 - 2l^2 p^4,\label{5He3}
\end{equation}
neglecting higher order terms. (It may be mentioned that some other
authors as noted in the introduction have ad hoc taken a third power
of $p$, and so on. However we should remember that these were all
phenomenological approaches.) Let us return to the Harmonic
oscillators in the background Dark Energy. The theory of these
Harmonic oscillators is well known (Cf. for example \cite{bd}). In
this usual theory we have a super position of Harmonic oscillator
solutions $(\hbar = 1 = c)$
$$f_k = e^{\imath (kx-\omega_k t)}$$
where we consider the one dimensional case, merely for simplicity
and
\begin{equation}
\omega_k = \sqrt{k^2+m^2}\label{a}
\end{equation}
Equation (\ref{a}) is a reflection of the usual energy momentum
relation
$$e^2 = p^2 + m^2$$
If now, we use instead the new dispersion relation (\ref{5He3})
above we will get, as can be easily verified,
$$\omega_k^2 = k^2 + m^2 - 2l^2 k^4$$
This shows that there is a reduction in the energy given by
\begin{equation}
k_{eff}^2 = k^2 - 2l^2 k^4\label{b}
\end{equation}
This is due to the appearance of a mass for the photon which would
not be there in the usual theory with (\ref{a}). Let us now estimate
this photon mass. As can be seen from (\ref{b}), we have,
\begin{equation}
m_\gamma = \left(\frac{\hbar}{c}\right) \left(k_{eff} -
k\right),\label{c}
\end{equation}
where we have restored $\hbar$ and $c$. If we consider $keV$
radiation as in the observations of Schaefer then we get for the
photon mass a value $\sim 10^{-65}gms$. If on the other hand we
consider $TeV$ or $GeV$ gamma rays, as are being observed then we
can easily deduce from (\ref{c}) that $m_\gamma \sim 10^{-62}gms$.\\
The important point is that latest observational estimates give an
improved upper limit for the photon mass $\sim 10^{-57}gms$
\cite{lake,it,tduniv}. Pleasingly our value is within this limit. It
may be mentioned that exactly this photon mass was deduced by the
author in the Planck oscillator - Dark Energy approach, quite
different from the approach given above \cite{bgsmp,tduniv}. Such a
photon mass can also be deduced on purely thermodynamic
considerations within the background Dark Energy \cite{bhtd,tduniv}.
Interestingly it was shown by Landsberg, using classical
thermodynamic theory that the above deduced photon mass is the
minimum allowable thermodynamic mass in the universe \cite{land}.
Exactly this mass was also proposed by Vigier and others
\cite{vigierieee} based on observational
evidence.\\
It can be easily seen that this photon mass (or equivalently the
above modified dispersion relation) leads to a dispersive velocity
for the photon \cite{vigierieee,tduniv}
\begin{equation}
v_\gamma = c \left[1 - \frac{m^2_\gamma c^4}{h^2
\nu^2}\right]^{1/2}\label{d}
\end{equation}
Equation (\ref{d}) shows the velocity dispersion with respect to
frequency though this is a very subtle effect which can be observed
in only Ultra High Energy Gamma Rays. Equation (\ref{d}) in fact
improves upon the limits of Schaefer and other authors, which were
used by Glashow and other authors, as mentioned in the introduction.
Moreover there have been claims that such a dispersive lag in the
arrival of High Energy Gamma Rays has already been observed
\cite{pav}. More recently Ellis and other authors have claimed such
a dispersive lag in the time arrival of Gamma Rays from an event in
the galaxy $mkn 537$ \cite{ellis}.
\section{Discussion}
1. We may mention that the photon having a mass does not really
contradict existing theory as pointed out by Deser \cite{deser}.\\
2. That the photon has a mass can also be deduced directly from the
background ZPF. Let us consider, following Wheeler a harmonic
oscillator in its ground state. The probability amplitude is
$$\psi (x) = \left(\frac{m\omega}{\pi \hbar}\right)^{1/4} e^{-(m\omega/2\hbar)x^2}$$
for displacement by the distance $x$ from its position of classical
equilibrium. So the oscillator fluctuates over an interval
$$\Delta x \sim (\hbar/m\omega)^{1/2}$$
The electromagnetic field for example is an infinite collection of
independent oscillators, with amplitudes $X_1,X_2$ etc. The
probability for the various oscillators to have amplitudes $X_1,
X_2$ and so on is the product of individual oscillator amplitudes:
$$\psi (X_1,X_2,\cdots ) = exp [-(X^2_1 + X^2_2 + \cdots)]$$
wherein there would be a suitable normalization factor. This
expression gives the probability amplitude $\psi$ for a
configuration $B (x,y,z)$ of the magnetic field that is described by
the Fourier coefficients $X_1,X_2,\cdots$ or directly in terms of
the magnetic field configuration itself by
$$\psi (B(x,y,z)) = P exp \left(-\int \int \frac{\bf{B}(x_1)\cdot \bf{B}(x_2)}{16\pi^3\hbar cr^2_{12}} d^3x_1 d^3x_2\right).$$
$P$ being a normalization factor. Let us consider a configuration
where the magnetic field or energy is everywhere zero except in a
region of dimension $l$, where it is of the order of $\sim \Delta
B$. The probability amplitude for this configuration would be
proportional to
$$\exp \left[-\left((\Delta B)^2 l^4/\hbar c\right)\right]$$
So the energy of fluctuation in a volume of length $l$ is given by
finally \cite{mwt,bgsfqv,bgscosfluc}
\begin{equation}
\mbox{Energy}\, \sim \frac{\hbar c}{l}\label{x}
\end{equation}
We can see from the above equation (\ref{x}) that this Energy in the
background ZPF is a minimum where $l$ is of the order of the radius
of the universe, that is $\sim 10^{28}cm$. Substitution in (\ref{x})
gives us back the above photon mass $m_\gamma \sim 10^{-65}gms$
\cite{tduniv}. As already pointed out this minimum mass agrees with
the minimum mass allowable in the universe from usual thermodynamic
theory, according
to Landsberg.\\
3. Another way of looking at this would be that the background Dark
Energy is a viscous medium with very small viscosity. This in fact
has been shown to lead back to the above mass \cite{bhtd,tduniv}.
Alternatively this can be seen in the following way \cite{kar}. The
Maxwell equations in a vacuum with a non zero conductivity
coefficient, can be shown to lead to a loss of energy of a photon
during its propagation. This is because the dissipating mechanism
leads to an extra term in the usual Maxwell's equations proportional
to
$$\frac{\partial}{\partial t} \vec{E}$$
This has been shown to lead to the non zero photon mass.\\
So a non zero photon mass was obtained based on a background Dark
Energy or ZPF (Cf. ref. \cite{tduniv}) at the Planck scale. On the
other hand using the Planck scale as a minimum scale, it is known
that spacetime geometry becomes noncommutative. This leads to a
dispersion relation which is a modification of the usual Lorentz
symmetry and equations like the Klein-Gordon and Dirac.\\
We have shown that, this alternative formulation leads to a photon
with a mass which is exactly of the same order, thus unifying both
the approaches. Finally not only is this photon mass well within the
experimental limits, but also leads to observable results in the
High Energy Gamma Ray spectrum. Hopefully NASA's GLAST satellite
will throw further light on this.
\section{The Modified Energy Momentum Formula for Fermions}
For Fermions the analysis can be more detailed, in terms of Wilson
lattices \cite{mont}. The free Hamiltonian now describes a
collection of harmonic fermionic oscillators in momentum space.
Assuming periodic boundary conditions in all three directions of a
cube of dimension $L^3$, the allowed momentum components are
\begin{equation}
{\bf q} \equiv \left\{q_k = \frac{2\pi}{L}v_k; k = 1,2,3 \right\},
\quad 0 \leq v_k \leq L - 1\label{4.59}
\end{equation}
(\ref{4.59}) finally leads to
\begin{equation}
E_{\bf q} = \pm \left(m^2 + \sum^{3}_{k=1} a^{-2} sin^2
q_k\right)^{1/2}\label{4.62}
\end{equation}
where $a = l$ is the length of the lattice, this being the desired
result leading to
$$E^2 = p^2e^2+m^2c^4 + \alpha l^2p^4$$
((\ref{4.62}) shows that $\alpha$ is positive.)
\section{A Modified Dirac Equation}
Once we consider a discrete spacetime structure, the energy momentum
relation, as noted, gets modified \cite{ijmpe2,mont} and we have in
units $c = 1 = \hbar$,
\begin{equation}
E^2 - p^2 - m^2 + l^2 p^4 = 0\label{6ce1}
\end{equation}
$l$ being the Planck length. Let us now consider the Dirac equation
\begin{equation}
\left\{ \gamma^\mu p_\mu - m\right\} \psi \equiv \left\{\gamma^\circ
p^\circ + \Gamma \right\} \psi = 0\label{6ce2}
\end{equation}
If we include the extra effect shown in (\ref{6ce1}) we get
\begin{equation}
\left(\gamma^\circ p^\circ + \Gamma + \beta l p^2\right) \psi =
0\label{6ce3}
\end{equation}
$\beta$ being a suitable matrix.\\
Multiplying (\ref{6ce3}) by the operator
$$\left(\gamma^\circ p^\circ - \Gamma - \beta l p^2\right)$$
on the left we get
\begin{equation}
p^2_0 - \left(\Gamma \Gamma + \left\{\Gamma \beta + \beta
\Gamma\right\} + \beta^2 l^2 p^4\right\} \psi = 0\label{6ce4}
\end{equation}
If (\ref{6ce4}), as in the usual theory, has to represent
(\ref{6ce1}), then we require that the matrix $\beta$ satisfy
\begin{equation}
\Gamma \beta + \beta \Gamma = 0, \quad \beta^2 = 1\label{6ce5}
\end{equation}
It follows that,
\begin{equation}
\beta = \gamma^5\label{6ce6}
\end{equation}
Using (\ref{6ce6}) in (\ref{6ce3}), the modified Dirac equation
finally becomes
\begin{equation}
\left\{\gamma^\circ p^\circ + \Gamma + \gamma^5 l p^2\right\} \psi =
0\label{6ce7}
\end{equation}
Owing to the fact that we have \cite{bd2}
\begin{equation}
P \gamma^5 = -\gamma^5 P\label{6ce8}
\end{equation}
It follows that the modified Dirac equation (\ref{6ce7}) is not
invariant under reflections.\\
We can also see that due to the modified Dirac equation
(\ref{6ce7}), there is no additional effect on the anomalous
gyromagnetic ratio. This is because, in the usual equation from
which the magnetic moment is determined \cite{merz} viz.,
$$\frac{d\vec{S}}{dt} = -\frac{e}{\mu c} \vec{B} \times \vec{S},$$
where $\vec{S} = \hbar \sum/2$ is the electron spin operator, there
is now an extra term
\begin{equation}
\left[\gamma^5, \sum\right]\label{6ce9}
\end{equation}
However the expression (\ref{6ce9}) vanishes by the property of the
Dirac matrices.\\
We would like to comment on the modified Dirac equation
(\ref{6ce7}). The modification is contained in the extra term
$\gamma^5 lp^2$. This essentially is an extra mass that shows up
that is the mass $m$ of the fermion becomes $m + \Delta m$. However
the curious feature is, that this extra term is, firstly independent
of the mass $m$, and secondly as in (\ref{6ce8}) is not invariant
under reflections.\\
If we now consider the case of a negligible mass, as in the two
component neutrino theory \cite{schweber}, (\ref{6ce7}) shows that a
supposedly massless particle acquires a mass, though this mass is
not reflection invariant. We can thus see an explanation for the non
zero neutrino mass.\\
Even if the fermion is massive, there is still the small correction
$\Delta m$ to its mass, though this again, according to (\ref{6ce7})
is not reflection invariant. It may be possible to detect this
subtle effect in very high energy collision perhaps even in the
context of $LHC$.\\
We finally observe that the term $\gamma lp^2$ corresponds to an
energy
$$E \sim 10^{21}eV$$
for an electron with a speed $c$, as can be easily calculated. More
realistically,
\begin{equation}
E \sim l^2 m^2 \theta^2 c^2\label{x}
\end{equation}
where $\theta c$ is the particle's speed and $m$ is its mass. For
ultra relativistic protons with $\theta < 1$, (\ref{x}) gives
$$E \sim \theta^2 \cdot 10^{27} eV$$
At this stage we comment on the above in a little detail, in the
context of the violation of Lorentz symmetry as seen in
(\ref{5He3}), for example. It has been suspected that Lorentz
symmetry is being violated from an observation of Ultra High Energy
Cosmic Rays (UHECR). In this case, given Lorentz symmetry there is
the GZK cut off such that particles above an energy of about
$10^{20}eV$ would not be able to travel cosmological distances and
reach the earth (Cf.
ref.\cite{uof,ijtp2,cu,mag2,jacobson,jacob2,amel} for
details).\\
So detection of cosmic rays arriving at the Earth with energies
above $10^{20}eV$ questions the presence of the $GZK$ cutoff
\cite{science1}. This cutoff determines the energy where the cosmic
ray spectrum is expected to abruptly drop according to a power law
in the energy. Cosmic rays with ultra high energies (above $\sim 5
\times 10^{19}eV$) lose energy through photoproduction of pions when
traveling through the cosmic microwave background radiation $(CMB)$.
An event of $10^{20}eV$ has to be produced within $\sim 100 Mpc$
unless there is non standard physics \cite{science2} and
\cite{science3}. So these events are a mystery,
the so called $GZK$ puzzle. Does this mean that Lorentz symmetry is being violated?\\
It is suspected that some twenty contra events have already been
detected, and phenomenological models of Lorentz symmetry violation
have been constructed by Glashow, Coleman and others while this also
follows from the author's fuzzy spacetime theory
\cite{gon,cole,jacob3,olin,carroll,nag}. The essential point here is
that the energy momentum relativistic formula is modified leading to
new effects.\\
What is very interesting is that already we are above the $GZK$
threshold.\\
Finally, we point out that using the modified dispersion relation
(\ref{5He3}), but for Fermions, for a massless particle, $m = 0$,
and identifying the extra term $l^2p^4$ as being due to a mass
$\delta m$, we can easily deduce that
$$\delta m = \frac{\hbar}{cl} \, \mbox{or}\, l =
\frac{\hbar}{c\delta m}$$ This shows that $l$ is the Compton
wavelength for this mass $\delta m$ or alternatively if $l$ is the
Compton wavelength, then we deduce the mass, now generated from the
extra effect.

\end{document}